\documentclass[cameraready]{Interspeech}

\title{Listen First, Then Answer: Timestamp-Grounded Speech Reasoning}

\usepackage{amsmath}
\usepackage[most]{tcolorbox}
\usepackage{xcolor}

\definecolor{royalblue}{rgb}{0.21,0.49,0.74}

\definecolor{lightpurple}{HTML}{9999FF}
\usepackage{multirow}
\usepackage[table]{xcolor}
\usepackage{amssymb}
\usepackage{pifont}
\usepackage{makecell}
\usepackage{booktabs}
\usepackage{makecell}
\usepackage{color}

\usepackage{cite}
\usepackage{graphicx}    
\usepackage{subcaption}  
\author[affiliation={1,2}]{Jihoon}{Jeong}
\author[affiliation={1,3}]{Pooneh}{Mousavi}
\author[affiliation={1,3}]{Mirco}{Ravanelli}
\author[affiliation={1,2}]{Cem}{Subakan}

\address{
    $^1$ Mila-Quebec AI Institute, $^2$ Université Laval, $^3$ Concordia University
}

\email{jhjeong@resl.kaist.ac.kr}

\keywords{Large Audio Language Models, Reasoning, Grounding, Interpretability}

\usepackage{comment}

\begin{document}

\maketitle

\begin{abstract}

Large audio-language models (LALMs) can generate reasoning chains for their predictions, but it remains unclear whether these reasoning chains remain grounded in the input audio. 
In this paper, we propose an RL-based strategy that grounds the reasoning outputs of LALMs with explicit timestamp annotations referring to relevant segments of the audio signal.
Our analysis shows that timestamp grounding leads the model to attend more strongly to audio tokens during reasoning generation.
Experiments on four speech-based benchmark datasets demonstrate that our approach improves performance compared to both zero-shot reasoning and fine-tuning without timestamp grounding. 
Additionally, grounding amplifies desirable reasoning behaviors, such as region exploration, audiology verification, and consistency, underscoring the importance of grounding mechanisms for faithful multimodal reasoning.

\end{abstract}

\begin{tcolorbox}[colback=gray!8, colframe=gray!50, arc=2mm, boxrule=0.5pt]
\small
\textbf{Project Page.}
For additional demo materials, please visit:
\href{https://ijihoon98.github.io/TGSR/}{\textcolor{royalblue}{ijihoon98.github.io/TGSR/}}.
\end{tcolorbox}    
\section{Introduction}
\label{sec:intro}

Large audio-language models (LALMs) \cite{chu2023qwen, kong2024audio, tang2023salmonn, gong2023joint, ghosh2024gama} have recently shown promising performance across a wide range of audio understanding and reasoning tasks. 
However, despite their multimodal design, most existing LALMs lack \textit{explicit mechanisms for grounding} their reasoning in concrete acoustic evidence.
As a result, model predictions are often driven by abstract linguistic priors rather than faithful engagement with the input audio signal.
This disconnect between reasoning and acoustic evidence can lead to fragile Chain-of-Thought (CoT) \cite{wei2022chain, wang2022self, kojima2022large} and compromise the faithfulness and interpretability of the explanations.

\begin{figure}[t]
    \centering
    \includegraphics[width=\linewidth]{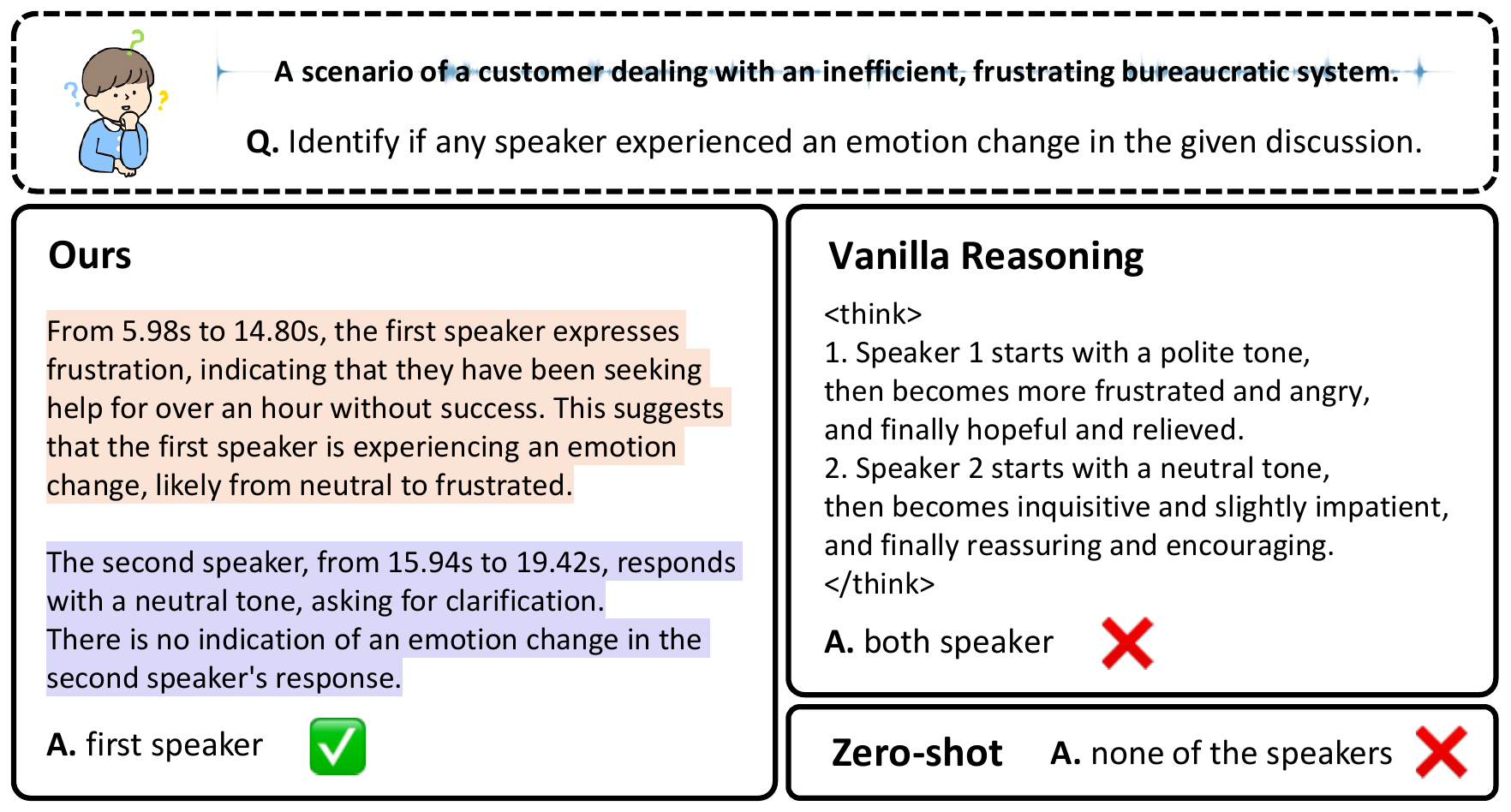}
    \caption{
    \textbf{Grounded timestamp reasoning enables interpretable and accurate answers.}
    Our method decomposes the task into a sequence of reasoning steps explicitly grounded in corresponding audio regions. In contrast, vanilla reasoning and zero-shot tend to produce ungrounded and incorrect responses.}
    \label{fig:teaser}
    \vspace{-2mm}
\end{figure}

In this paper, we hypothesize that LALMs both “hear better” and “reason better” when their textual reasoning steps are explicitly grounded in precise temporal segments of the audio signal. 
This hypothesis is inspired by human auditory cognition, where listeners dynamically shift their attentional focus over time to selectively attend to task-relevant acoustic events during reasoning and decision-making \cite{yang2016theoretical, friston2021active}.
As shown in Fig. \ref{fig:teaser}, timestamp-based grounding promotes more targeted and systematic cross-referencing between linguistic reasoning and acoustic evidence throughout multi-step inference. 
Hence, we utilize text-to-timestamp object-centric grounding as the intermediate reasoning stage, where the predicted timestamp zones with explanation serve as simple but effective CoT signals to help improve the quality of the final reasoning step.

Our contributions can be summarized as follows:
\begin{itemize}
    \item We identify and analyze a fundamental limitation of current LALMs, showing that correct predictions do not necessarily imply faithful audio grounding, and that existing models often rely on text-biased reasoning with weak engagement of acoustic evidence.
    
    \item We propose a timestamp-grounded reasoning framework that anchors reasoning steps to temporally localized audio segments, with a two-stage strategy that combines supervised timestamp alignment and reward-based optimization.
    
    \item We present comprehensive evaluations across multiple speech benchmarks and behavior-level analyses, demonstrating that timestamp grounding improves not only end-task accuracy but also reasoning consistency, audio verification behavior, and overall interpretability.
\end{itemize}

\section{Related Works}
\label{sec:related_works}

\begin{figure*}[t]
    \centering
    \resizebox{0.95\linewidth}{!}{
    \includegraphics[width=\linewidth]{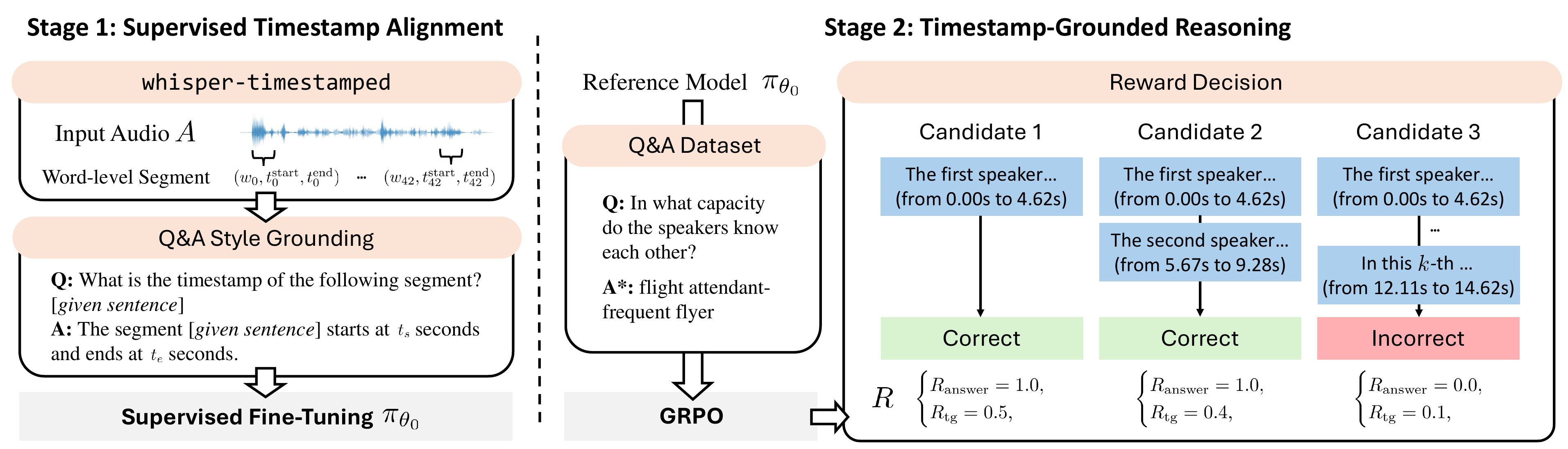}
    }
    \vspace{-2mm}
    \caption{
        \textbf{Overview of the proposed framework.}
        Stage 1 learns temporal localization through supervised timestamp alignment. Stage 2 performs GRPO-based optimization to encourage answers justified by concise timestamp-grounded reasoning.
    }
    \label{fig:main_figure}
    \vspace{-2mm}
\end{figure*}

\textbf{Reasoning in Audio-Language Models.} 
CoT reasoning has emerged as an effective paradigm for improving the transparency and accuracy of large language models by modeling intermediate reasoning steps \cite{jaech2024openai, guo2025deepseek, team2025kimi, wang2025they}. 
This paradigm has been extended to LALMs, where embedded reasoning integrates linguistic abstractions, including models such as the Thinker variants of the Qwen family \cite{xu2025qwen3, yang2025qwen3} and Audio Flamingo 3 \cite{goel2025audio}.
Recent studies aim to enhance reasoning quality, showing that it can be achieved even with lightweight models \cite{deshmukh2025mellow}, while works such as Audio-CoT \cite{ma2025audio} and Audio-Reasoner \cite{xie2025audio} further investigate explicit and structured CoT pipelines via standard fine-tuning.
Moreover, reinforcement learning has been adopted to further enhance reasoning fidelity, including R1-AQA \cite{li2025reinforcement}, Omni-R1 \cite{rouditchenko2025omni}, and Audio-Thinker \cite{wu2025audio}, which fine-tune open-weight audio models using Group Relative Policy Optimization \cite{shao2024deepseekmath}.
While these approaches achieve notable gains in reasoning accuracy, they primarily \textit{emphasize performance improvements rather than explicitly assessing whether reasoning steps are grounded in actual audio evidence.}


\textbf{Grounded Reasoning for Multimodal Language Models.}
Recent studies have explored grounded reasoning as a means to promote cross-referencing between textual and non-textual modalities.
In the vision-language domain, prior works such as Visual-RFT \cite{liu2025visual}, MM-GCoT \cite{wu2025grounded}, ARGUS \cite{man2025argus}, and ViGoRL \cite{sarch2025grounded}  integrate visual signals, most commonly object bounding boxes, into the reasoning process, thereby improving faithfulness between reasoning steps and perceptual evidence. 
In contrast, grounded reasoning in LALMs remains largely underexplored. 
Step-Audio-R1 \cite{tian2025step} represents an initial attempt to introduce audio grounding into reasoning; however, its grounding is not quantitatively measured, and the presence of acoustic references is determined via heuristic criteria. 
Relatedly, Timestamped Audio Captioner (TAC) \cite{kumar2026tac} proposes temporally grounded audio captioning, generating time-aligned descriptions. 
However, TAC itself is not a reasoning model; rather, it serves as a captioning system whose outputs are provided as inputs to downstream LLMs.
By contrast, our work directly addresses these limitations \textit{by internally assessing whether real audio evidence is utilized, and by strengthening the explainability of CoT through auditory interactions.}

\section{Methodology}
\label{sec:methodology}

Our goal is to equip the reasoning chains generated by LALMs with timestamps that correspond to salient segments of the input audio. 
To this end, we propose a two-stage training framework that progressively builds robust timestamp grounding capabilities within the reasoning.
In the first stage, the model is trained solely on timestamp prediction; in the second stage, timestamp prediction is integrated into downstream QA reasoning tasks.
The overall structure is represented in Fig. \ref{fig:main_figure}.


\subsection{Stage 1: Supervised Timestamp Alignment}
\label{subsec:supervised_timestamp_alignment}
A key obstacle to faithful audio reasoning is that most existing LALMs are not reliably capable of extracting temporally precise evidence. \cite{kumar2026tac}
In practice, timestamp references are often missing, coarse, or misaligned with the true acoustic speech events, which can propagate to erroneous CoT.
Therefore, before optimizing high-level reasoning behaviors, we first establish a strong temporal grounding primitive through supervised timestamp alignment (STA).

\noindent \textbf{Timestamped supervision.}
To construct timestamp supervision at scale, we build a timestamp-annotated speech corpus using \texttt{whisper-timestamped} \cite{lintoai2023whispertimestamped}, an extension of Whisper \cite{radford2023robust} that produces word-level timestamps and confidence scores.
Unlike vanilla Whisper, which primarily provides segment-level timestamps, \texttt{whisper-timestamped} estimates word timestamps by applying Dynamic Time Warping \cite{JSSv031i07} to cross-attention weights, enabling finer and more reliable temporal localization.

\noindent \textbf{From timestamped transcript to Q\&A-style grounding data.}
Given an audio file $A$, we first obtain a word-level timestamped transcript $\{(w_j, t^{\text{start}}_j, t^{\text{end}}_j)\}_j$.
We then convert the timestamped transcript into question-answer instances designed to train explicit temporal anchoring, as shown in Fig. \ref{fig:main_figure}.
Specifically, we sample a text phrase from the transcript as a query and ask the model to predict the corresponding temporal region in input audio $A$, where [\emph{given sentence}] indicates a spoken sentence whose start ($t_s$) and end ($t_e$) times are to be grounded.
This method trains the model to generate start-end timestamps, making it compatible with our timestamp-grounded reasoning framework and producing a reference model, denoted as $\pi_{\theta_0}$.

\subsection{Stage 2: Timestamp-Grounded Reasoning}
\label{subsec:timestamp_grounded_reasoning}
With an established textual reasoning foundation, we now address the core challenge of equipping the model’s reasoning process with explicit timestamp grounding. 
We adopt \emph{Group Relative Policy Optimization (GRPO)} \cite{shao2024deepseekmath}, a reinforcement learning framework that stabilizes policy learning by leveraging relative comparisons among multiple sampled trajectories.

\noindent \textbf{Answer Correctness Reward ($R_{\text{answer}}$).}
The first component, $R_{\text{answer}}$, measures whether the model produces the correct multiple-choice answer.
We extract the predicted option from the final line and assign a binary reward:
\begin{equation}
R_{\text{answer}} =
\begin{cases}
1.0, \quad \text{if pred. matches GT}. \\
0.0, \quad \text{otherwise}.
\end{cases}
\end{equation}
This term ensures that reward-based refinement preserves task-level accuracy and encourages the generated reasoning to remain grounded in the final answer.

\noindent \textbf{Timestamp Grounded Reward ($R_{\text{tg}}$).}
While $R_{\text{answer}}$ evaluates \emph{what} answer the model produces, it does not assess \emph{how} the answer is justified.
To promote concise and meaningful timestamp-grounded reasoning, we introduce a timestamp grounding reward $R_{\text{tg}}$, which evaluates whether the generated reasoning refers to relevant audio timestamps and encourages compact timestamp usage.
This design is aligned with prior works \cite{wu2025audio, dumitru2025conciserl, yue2025don} on effective reasoning that emphasizes reducing unnecessary steps while preserving semantic evidence.


We identify distinct grounding units that explicitly reference temporal segments of the input audio.
Let $k$ denote the \emph{number of timestamp-grounded units} in the generated completion.
The compaction score $R_{\text{tg}}$ is defined as:
\begin{equation}
R_{\text{tg}} =
\begin{cases}
0,  k = 0; \quad C_{\max},  k \le k_{\text{ref}}; \quad C_{\min},  k \ge k_{\text{max}}; \\
C_{\max} - (k - k_{\text{ref}})
\frac{C_{\max} - C_{\min}}{k_{\text{max}} - k_{\text{ref}}}, \quad \text{otherwise}.
\end{cases}
\end{equation}
where $k_{\text{ref}}$ denotes the reference number of timestamped reasoning units (set to $1$ in our experiments), and $C_{\max}$ and $C_{\min}$ denote the maximum and minimum compaction rewards (set to $0.5$ and $0.1$ respectively).
This formulation encourages concise yet sufficiently grounded reasoning, avoiding both missing grounding signals ($k=0$) and overly verbose timestamp usage.



\noindent \textbf{Final Reward.}
The final reward is computed as:
\begin{equation}
R = R_{\text{answer}} + R_{\text{tg}}
\end{equation}

This formulation encourages the model to not only predict the correct answer, but also justify it using concise, temporally grounded evidence, as shown in Fig. \ref{fig:teaser}, \ref{fig:main_figure}.

\section{Experimental Settings}
\label{sec:experiments}

\subsection{Datasets}
\label{subsec:datasets}
We use a mixture of timestamp-annotated speech datasets for Stage~1 training, including LibriSpeech \cite{panayotov2015librispeech}, CoVoST~2 \cite{wang2020covost}, MELD \cite{poria2019meld}, multi-speaker \cite{xie2025audio}, and YouTube8M \cite{abu2016youtube} speech data. 
The resulting corpus contains approximately 268k examples, which we plan to release to facilitate future research.
For Stage~2, we perform reward-based optimization using question-answering datasets, including MELD \cite{poria2019meld}, multi-speaker dialogue QA \cite{xie2025audio}, and YouTube8M speech QA \cite{abu2016youtube}, resulting in approximately 47k training examples.

\subsection{Implementation Details}
\label{subsec:implementation}
We conduct our experiments using Qwen2.5-Omni \cite{xu2025qwen25omni} and Audio Flamingo 3 \cite{goel2025audio} as the base models. Training is performed on a single node with 8 H200 GPUs. 
We use a per-GPU batch size of 8 with 4 gradient accumulation steps. The learning rate is set to $2\times10^{-5}$ for Stage 1, $5\times10^{-6}$ for Stage 2, with a sampling temperature of 0.8, 8 responses per GRPO step, and a KL regularization coefficient $\beta=0.04$.

\begin{figure}[t]
    \centering
    \resizebox{0.80\linewidth}{!}{
    \includegraphics[width=\linewidth]{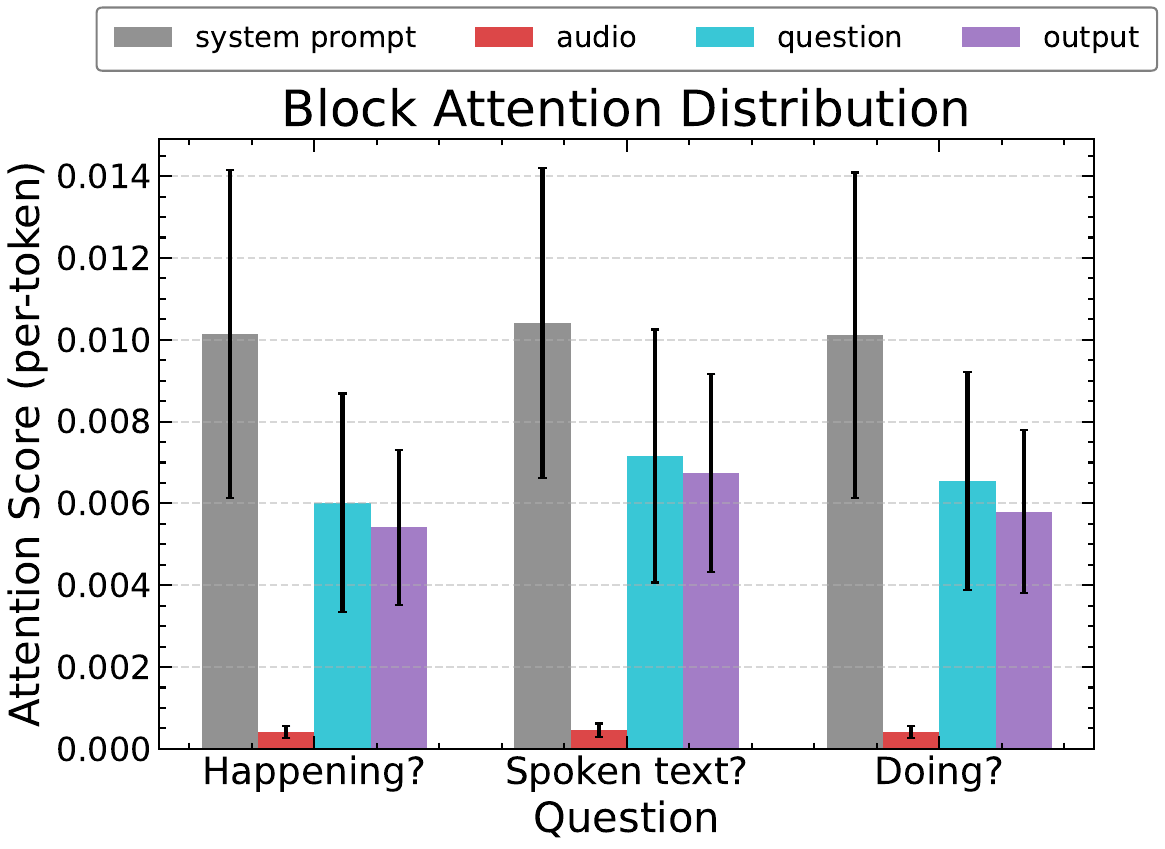}
    }
    \vspace{-2mm}
    \caption{
    \textbf{Semantic attention distribution of a baseline LALM across input categories.}
    It aggregates token-level attention by semantic role, highlighting under-attention to audio features.}
    \label{fig:audio_attention_blocks}
    \vspace{-4mm}
\end{figure}



\section{Results}
\label{sec:experiments}



\subsection{On the Limited Use of Audio Information in LALMs}
\label{subsec:limited_audio_info}

Despite their multimodal design, we observe that current LALMs exhibit a pronounced text-biased reasoning tendency, closely resembling previously reported in vision-language models \cite{pezeshkpour2025mixed, vo2025vision}. 
To analyze this behavior more systematically, we conduct a \textbf{semantic-based attention analysis} that aggregates attention weights over system tokens, audio tokens, instruction tokens, and self-referential outputs.
We perform this analysis on the MMAU \cite{sakshi2024mmau} speech data using audio-critical queries, \{\emph{What is happening?}, \emph{What is the spoken text?}, and \emph{What are they doing?}\}, for which the answer cannot be inferred from textual instructions alone and requires access to the audio signal.


As shown in Fig.~\ref{fig:audio_attention_blocks}, baseline LALMs allocate only a small fraction of attention to audio tokens, even when producing correct or seemingly audio-grounded answers. 
We observe a clear attention sink effect: system tokens dominate attention allocation even on a per-token basis, receiving over 15$\times$ more attention than audio tokens.
These findings indicate that successful predictions do not necessarily imply faithful audio reference. 
As a result, models may generate plausible yet weakly grounded explanations that do not reflect the true causal role of audio in the reasoning process.
This behavior motivates the need for training strategies that explicitly encourage alignment between reasoning steps and grounded audio evidence.

\subsection{Grounding is Listening}
\label{subsec:grounding_is_listening}

\begin{table}[t]
\centering
\caption{Effect of Sentence-level Timestamp Alignment.}
\vspace{-2mm}
\label{tab:timestamp_iou}
\resizebox{0.95\linewidth}{!}{
\begin{tabular}{l c c c c}
\toprule
\textbf{Model} & \textbf{STA} & \textbf{IoU $\uparrow$} & \textbf{F1 $\uparrow$} & \textbf{IoU $\geq$ 0.7 (\%) $\uparrow$} \\
\midrule
\multirow{2}{*}{Qwen-2.5-Omni \cite{xu2025qwen25omni}}
  & \ding{55} & 0.2324 & 0.3074 & 11.32 \\
  & $\checkmark$ & {0.7189} & {0.7675} & {69.79} \\
\midrule
\multirow{2}{*}{AudioFlamingo3 \cite{goel2025audio}}
  & \ding{55} & 0.2838 & 0.3592 & 11.45 \\
  & $\checkmark$ & 0.5439 & 0.6012 & 47.95 \\
\midrule
\midrule
Gemini-2.5-Flash \cite{comanici2025gemini}
  & -- & 0.7356 & 0.8141 & 71.08 \\
\bottomrule
\end{tabular}
}
\end{table}

\begin{figure}[t]
    \centering
    \includegraphics[width=\linewidth]{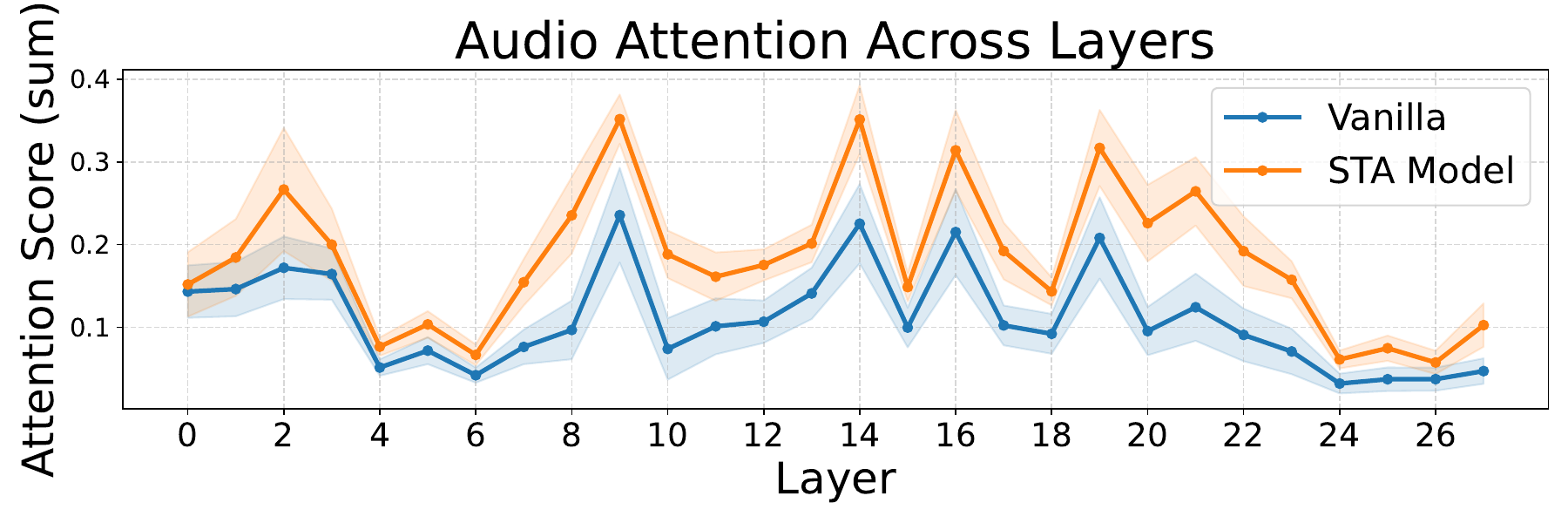}
    \vspace{-4mm}
    \caption{
    \textbf{Layer-wise audio attention during timestamp generation.}
    STA model allocates higher attention to audio tokens across transformer layers compared to the vanilla model, indicating stronger reliance on acoustic evidence.
    }
    \label{fig:audio_attention_layers}
    \vspace{-4mm}
\end{figure}



In this section, we analyze \emph{whether LALMs genuinely rely on audio evidence during grounding} and evaluate the effectiveness of timestamp supervision.
Before we dive in, STA improves timestamp grounding accuracy, as reflected by consistent gains in Table~\ref{tab:timestamp_iou}, while the vanilla model shows timestamp hallucinations.
We report the intersection-over-union (IoU) between predicted timestamp ranges and ground-truth segments, the proportion of predictions achieving high-overlap alignment (IoU $\geq$ 0.7), and Sound Event Detection F1 score (SED-F1) \cite{mesaros2016metrics}, which measures event-level temporal precision. 
Across all metrics, our method outperforms the vanilla baseline and achieves performance comparable to Gemini-2.5-flash \cite{comanici2025gemini}, indicating robust temporal localization.

Next, Fig.~\ref{fig:audio_attention_layers} provides empirical evidence supporting the claim that \textbf{grounding is listening}.
We compare semantic audio attention maps between the vanilla and the timestamp-aligned model, focusing specifically on the \emph{time generation} phase, where the model is required to produce explicit temporal references.
The timestamp-aligned model allocates higher attention to audio tokens across layers during time generation, suggesting that improved grounding is closely tied to faithful listening to temporally relevant audio cues.


Notably, we observe that certain layers exhibit a pronounced shift in listening behavior, acting as effective \textit{picking points} where acoustic evidence is selectively amplified.
This observation is consistent with prior analyses of the LLM domain \cite{lindsey2025biology, skean2025layer}, which indicate that different layers internalize distinct functional roles.
From this perspective, STA appears to reshape the behavior of layers that integrate external evidence, leading to more faithful incorporation of audio cues into the model’s internal representations.

\begin{table}[t]
\centering
\caption{Performance comparison on MMAU-mini-Speech, MMAR-Speech, AIR-Bench, and MELD.}
\small
\setlength{\tabcolsep}{6pt}
\resizebox{1.0\linewidth}{!}{
\begin{tabular}{l c c c c c c c}
\toprule
\multirow{2}{*}{\textbf{Methods}} & \multirow{2}{*}{\textbf{Size}} 
& \multicolumn{1}{c}{\textbf{MMAU-mini}} 
& \multicolumn{1}{c}{\textbf{MMAR}} 
& \multicolumn{3}{c}{\textbf{AIR-Bench}} 
& \multicolumn{1}{c}{\textbf{MELD}} \\
\cmidrule(lr){3-3} \cmidrule(lr){4-4} \cmidrule(lr){5-7} \cmidrule(lr){8-8}
 &  & Speech (\%) & Speech (\%) & SER (\%) & SNV (\%) & SIC (\%) & (\%) \\
\midrule

\rowcolor{gray!10}
\multicolumn{8}{l}{\textbf{\textit{Proprietary Models}}} \\
Gemini 2.5 Flash \cite{comanici2025gemini} & -- & \textbf{75.08} & \textbf{72.11} & 56.4 & 68.5 & \underline{88.6} & 61.5 \\
GPT-4o Audio \cite{hurst2024gpt}  & -- & 66.67 & \underline{70.41} & 51.2 & 61.6 & \textbf{89.3} & 62.5 \\
\midrule

\rowcolor{gray!10}
\multicolumn{8}{l}{\textbf{\textit{Open-source Models}}} \\
SALMONN \cite{tang2023salmonn}         & 7B & 26.43 & 24.35 & 29.9 & 34.3 & 42.3 & 37.2 \\
Audio Flamingo 3 \cite{goel2025audio} & 7B & 66.37 & 57.48 & 59.5 & \textbf{76.8} & 79.6 & 58.5 \\
\midrule

\rowcolor{gray!10}
\multicolumn{8}{l}{\textbf{\textit{Audio Reasoning Methods}}} \\
Audio-CoT \cite{ma2025audio}       & 8.4B & 55.26 & 34.01 & -- & -- & -- & -- \\
Audio-Reasoner \cite{xie2025audio}  & 8.4B & 66.07 & 32.99 & \underline{60.5} & 56.3 & 88.1 & 63.2 \\
Audio-Thinker \cite{wu2025audio}   & 8.4B & 73.37 & 64.29 & 56.2 & 67.5 & -- & -- \\
\midrule

\rowcolor{gray!10}
\multicolumn{8}{l}{\textbf{\textit{Ablation Variants}}} \\
Qwen2.5-omni \cite{xu2025qwen25omni}    & 7B & 70.60 & 59.86 & 60.2 & 63.9 & 83.5 & 60.3 \\
\setlength{\parindent}{1em}+ Only STA
                 & 7B & 71.17 & 61.22 & 59.5 & 66.0 & 84.3 & 62.8 \\
\setlength{\parindent}{1em}+ Reasoning SFT
                 & 7B & \underline{74.47} & 62.93 & 58.5 & 68.1 & 85.0 & 61.8 \\
\textbf{Ours}    & \textbf{7B} & \underline{74.47} & 64.63 & \textbf{62.5} & \underline{70.4} & \textbf{89.3} & \textbf{64.6} \\
\bottomrule
\end{tabular}
}
\label{tab:benchmark_comparison}
\vspace{-4mm}
\end{table}

\subsection{Timestamp Grounding Improves Benchmark Ability}
\label{subsec:grounding_benchmark}

Table~\ref{tab:benchmark_comparison} compares our method with \textbf{\emph{Proprietary}} and \textbf{\emph{Open-source}} LALMs on four representative benchmarks: MMAU-mini-Speech \cite{sakshi2024mmau}, MMAR-Speech \cite{ma2025mmar}, AIR-Bench \cite{yang2024air} and MELD \cite{poria2019meld}.
These benchmarks evaluate speech understanding, speech reasoning, and dialog-level recognition, providing a comprehensive assessment of speech-centric ability.

Starting from the Qwen2.5-Omni \cite{xu2025qwen25omni} baseline, we first analyze the impact of individual components through an \textbf{\emph{Ablation Variants}}. 
Applying only STA shows minor improvements, suggesting that it does not fully resolve domain alignment. 
We further evaluate a reasoning SFT variant \emph{without timestamp supervision}, which improves over the baseline on several benchmarks but underperforms our model. 
These results indicate that timestamp alignment and reasoning supervision alone are each insufficient, highlighting the need to integrate temporal grounding with reasoning to achieve robust performance.

Our full model achieves the best overall performance across all benchmarks. 
In particular, it obtains the strongest results on AIR-Bench and MELD, even if compare with \textbf{\emph{Proprietary Models}}, demonstrating robust speech understanding and integration of acoustic cues. 
Moreover, on speech reasoning benchmarks such as MMAU-mini-Speech and MMAR-Speech, our method outperforms existing \textbf{\emph{Audio Reasoning Methods}}. 
Despite being designed for reasoning, prior audio reasoning methods show limited improvements on these benchmarks, whereas our model achieves stronger performance.
This observation indicates that grounding-based supervision provides a more effective foundation for speech reasoning.

\begin{table}[t]
\centering
\caption{Average visual reasoning behaviors per example on MMAU-mini-Speech.}
\vspace{-2mm}
\label{tab:behavior_analysis}
\small
\setlength{\tabcolsep}{6pt}
\resizebox{0.9\linewidth}{!}{
\begin{tabular}{l c c c c c}
\toprule
\textbf{Model} 
& \makecell{\textbf{Regions} \\ \textbf{Explored}}
& \makecell{\textbf{Audiology} \\ \textbf{Verify $\uparrow$}}
& \textbf{Consistency $\uparrow$}
& \textbf{Acc. $\uparrow$} \\
\midrule

\rowcolor{gray!12}
\multicolumn{5}{l}{\textbf{\textit{Zero-Shot}}} \\
Standard CoT 
& 1.3 & 0.27 & 0.72 & 69.4 \\
\rowcolor{gray!12}
\multicolumn{5}{l}{\textbf{\textit{RL-tuned}}} \\
w/o compaction 
& 3.6 & \underline{0.48} & \underline{0.78} & 72.8 \\
w/o grounding 
& 0.5 & 0.22 & 0.69 & \underline{74.2} \\
\textbf{Ours} 
& 1.8 & \textbf{0.56} & \textbf{0.83} & \textbf{74.5} \\

\bottomrule
\end{tabular}
}
\vspace{-2mm}
\end{table}

\begin{table}[t]
\centering
\caption{Average audio attention behaviors when generating reasoning on MMAU-mini-Speech.}
\vspace{-2mm}
\label{tab:attention_behavior_analysis}
\small
\setlength{\tabcolsep}{6pt}
\resizebox{0.85\linewidth}{!}{
\begin{tabular}{l c c c c}
\toprule
\textbf{Model} 
& \textbf{Vanilla}
& \makecell{\textbf{Only} \\ \textbf{STA}}
& \makecell{\textbf{Reasoning} \\ \textbf{SFT}}
& \textbf{Ours} \\
\midrule

\textbf{Audio Attention (All)}
& 0.0748
& 0.0936
& 0.0975
& \textbf{0.1030} \\

\textbf{Reasoning Attention}
& 0.0790
& 0.0952
& 0.1032
& \textbf{0.1138} \\

\textbf{Answer Attention}
& 0.0421
& 0.0690
& \textbf{0.0714}
& 0.0680 \\

\bottomrule
\end{tabular}
}
\vspace{-4mm}
\end{table}

\subsection{Timestamp Grounding Improves Behavior Quality}
\label{subsec:grounding_behavior}

Table~\ref{tab:behavior_analysis} goes beyond end-task accuracy and examines how timestamp grounding reshapes the model’s reasoning behavior.
\emph{Regions Explored} measures the number of distinct timestamped reasoning segments produced by the model.
\emph{Audiology Verify} evaluates the alignment between timestamped explanations and their corresponding audio segments using a Whisper-based similarity score \cite{radford2023robust}.
Finally, \emph{Consistency} measures the logical alignment between the model’s reasoning trace and its final prediction, evaluated by a judge model (Qwen3-32B \cite{yang2025qwen3}).

Compared to the zero-shot baseline, \textbf{\emph{RL-tuned variants}} show noticeable changes.
The model w/o compaction explores more regions and increases audio verification, but achieves lower accuracy.
In contrast, the variant w/o grounding achieves reasonable accuracy yet shows minimal region exploration and weak audio verification, suggesting that objective-driven training encourages shortcut strategies.
Our full model achieves the highest audio verification and the strongest reasoning consistency, although it explores fewer regions.

Table~\ref{tab:attention_behavior_analysis} reports the average audio attention allocated to audio tokens across the \textbf{\emph{Ablation Variants}} introduced in Table~\ref{tab:benchmark_comparison}.
Compared to the vanilla zero-shot reasoning baseline, all ablation variants allocate more attention to audio inputs during reasoning, suggesting that grounding signals and reasoning supervision encourage the model to rely more on acoustic evidence when forming reasoning chains. 
Despite this general increase, our full model achieves the highest audio attention overall, with the most pronounced gains observed during the reasoning stage.
These results indicate that timestamp-grounded reasoning encourages the model to verify auditory evidence, moving toward mechanistic interpretability in LALMs.
\section{Conclusion}
\label{sec:conclusion}

In this work, we show that correct predictions in LALMs do not necessarily imply faithful grounding in acoustic evidence. 
To address this, we propose a timestamp-grounded reasoning framework that anchors intermediate reasoning steps to temporally localized audio segments, allowing both the model and users to access the acoustic evidence supporting each reasoning step.
Experiments demonstrate that timestamp grounding improves benchmark performance, reasoning consistency, audio verification behavior, and interpretability. 
While our experiments focus on speech-centric tasks, out methodology is not conceptually limited to speech, and extending the framework to non-speech audio events is a plausible direction for future work.


\bibliographystyle{IEEEtran}
\bibliography{references}

\newpage
\appendix
\onecolumn

\begin{center}
    {\Large \textbf{Listen First, Then Answer: Timestamp-Grounded Speech Reasoning}}\\
    \vspace{0.5em}{\Large Appendix} \\
    \vspace{0.5em}
    \vspace{1.em}
\end{center}



\noindent
The structure of this Appendix is as follows:

\begin{itemize}
\item \textbf{Appendix \ref{section:A}} - Attention Map Analysis Details.
\item \textbf{Appendix \ref{section:B}} - Additional Experimental Settings.
\item \textbf{Appendix \ref{section:C}} - Additional Methodological Details.
\item \textbf{Appendix \ref{section:D}} - Limitations and Future Work.
\item \textbf{Appendix \ref{section:E}} - Qualitative Examples.
\end{itemize}




\section{Attention Map Analysis Details}
\label{section:A}

\subsection{Layer-wise Aggregated Attention Analysis Across Query Types}
\label{sec:appendix_layer_attention_queries}

\begin{figure}[h]
    \centering
    \resizebox{0.8\linewidth}{!}{
    \includegraphics[width=\linewidth]{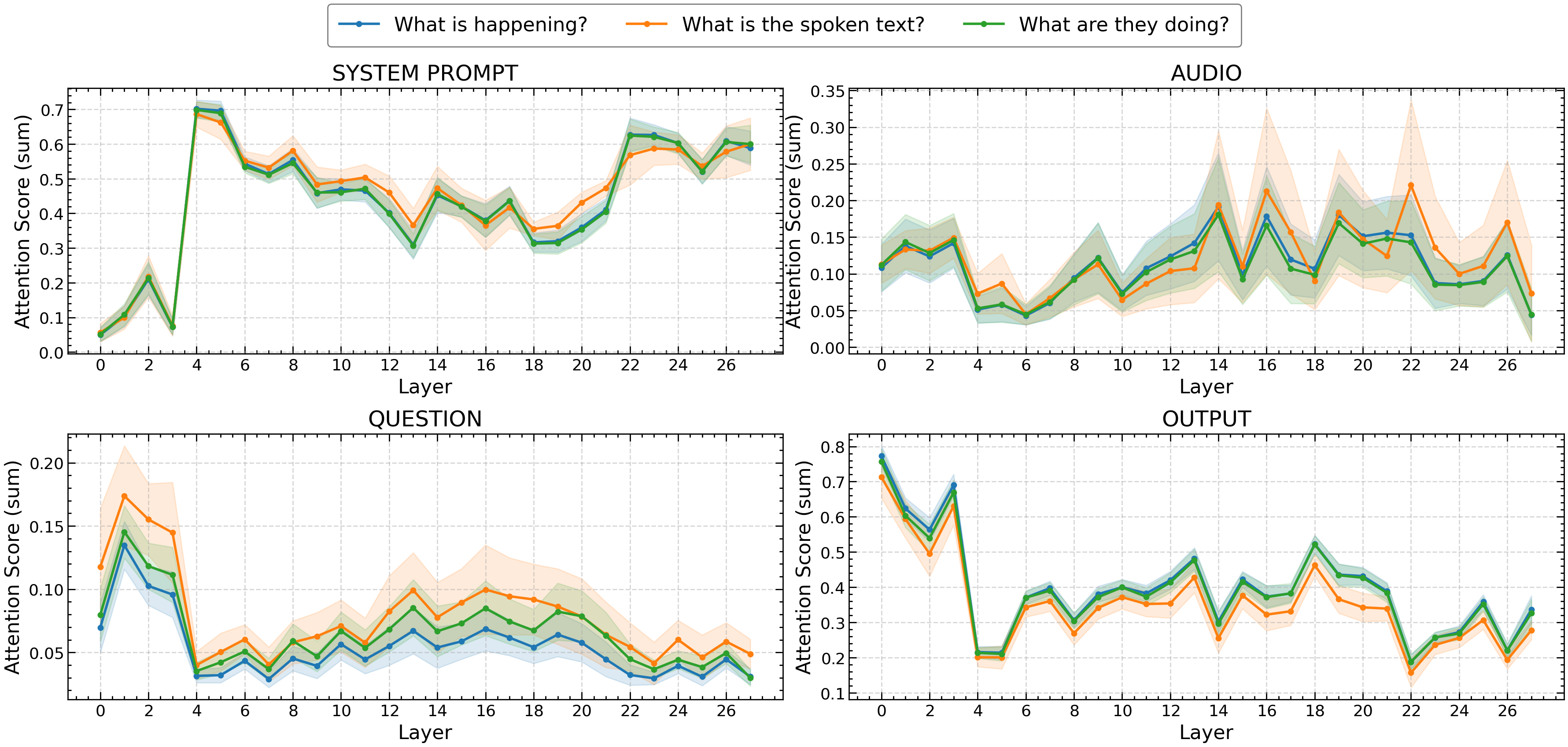}
    }
    \caption{
    \textbf{Layer-wise attention aggregation across different query types.}
    Unlike Fig.~\ref{fig:audio_attention_blocks}, the values shown here correspond to summed attention scores within each semantic block rather than per-token normalized values.
    The overall attention profiles remain highly consistent across layers, indicating that LALMs follow a similar internal processing pattern for audio-conditioned reasoning tasks.
    }
    \label{fig:open_question_audio_attention_per_outputs}
\end{figure}

To further analyze the internal attention behavior of LALMs, we present a layer-wise attention analysis aggregated over semantic token groups.
Unlike Fig.~\ref{fig:audio_attention_blocks} in the main paper, which reports attention values normalized by the number of tokens in each semantic block, the results shown here correspond to the \textbf{summed attention scores} within each block.
This complementary view allows us to verify that the observed attention patterns are not artifacts of token normalization.

We perform this analysis using three different audio-critical queries reported in Section~\ref{subsec:limited_audio_info}:
\emph{What is happening?}, \emph{What is the spoken text?}, and \emph{What are they doing?}.
These queries require access to the acoustic signal and cannot be answered from textual instructions alone.
As shown in Fig.~\ref{fig:open_question_audio_attention_per_outputs}, the layer-wise attention patterns remain highly consistent across the three query types.
In particular, system tokens dominate the attention allocation across layers, while the attention assigned to audio tokens remains relatively small.
This observation is consistent with the attention sink effect reported in Section~\ref{subsec:limited_audio_info}.
Interestingly, the similarity of attention profiles across different query formulations suggests that LALMs process audio-conditioned reasoning in a relatively uniform manner at the architectural level.
In other words, the internal attention dynamics appear to follow a stable processing pattern regardless of the specific wording of the query.
These results provide additional evidence that current LALMs rely heavily on textual context while under-utilizing acoustic information, motivating the need for training strategies that explicitly encourage stronger alignment between reasoning and audio evidence.

\subsection{Case Study: Attention Aggregation during Timestamp Generation}
\label{sec:appendix_attention_case}

To provide an intuitive understanding of how timestamp grounding emerges in our framework, we present a qualitative case study illustrating the attention aggregation behavior during timestamp generation.

\vspace{2mm}

\textbf{Example Build-up.} Given the following query:

\begin{quote}
\textbf{Q.} What is the timestamp of the following segment?  
``I hope the scientist who confirms stream theory,''
\end{quote}

the ground-truth timestamp is:

\begin{quote}
\textbf{Answer:} (12.05 -- 14.29 s)
\end{quote}

However, a zero-shot model without timestamp alignment produces the following prediction:

\begin{quote}
\textbf{Zero-shot prediction:}  
``The segment `I hope the scientist who confirms stream theory' starts at 18.96 seconds and ends at 21.6 seconds.''
\end{quote}

In contrast, after Stage 1 training with supervised timestamp alignment (STA), the model produces a prediction that closely matches the ground truth:

\begin{quote}
\textbf{After STA (Stage 1):}  
``The segment `I hope the scientist who confirms stream theory,' starts at 12.02 seconds and ends at 14.34 seconds.''
\end{quote}

\begin{figure}[h]
    \centering
    \resizebox{0.75\linewidth}{!}{
    \includegraphics[width=\linewidth]{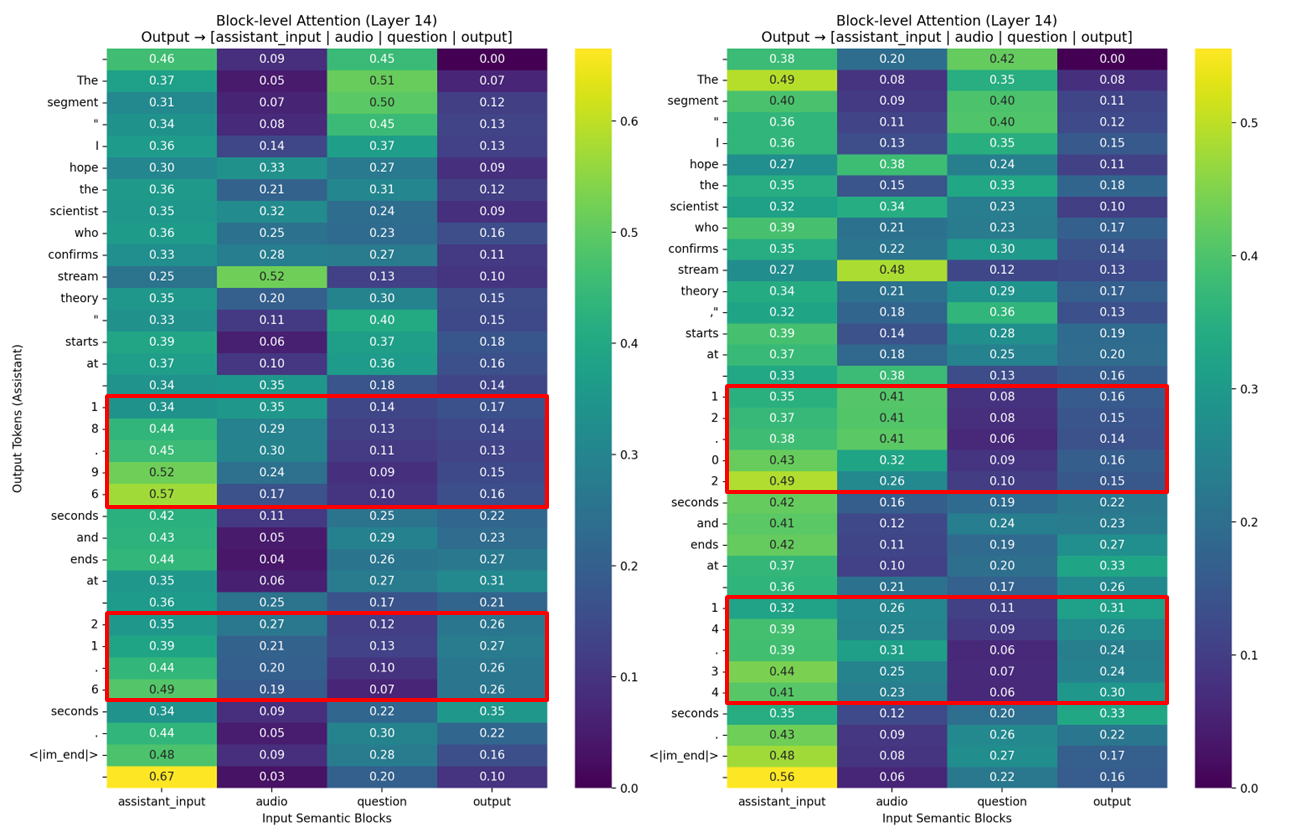}
    }
    \caption{
    \textbf{Block-level attention aggregation at Layer 14 during timestamp generation.}
    Attention weights from each output token are aggregated across four semantic input blocks.
    Tokens corresponding to timestamp prediction (highlighted in a red box) allocate increased attention to the audio block, indicating that temporal reasoning relies on acoustic evidence.
    }
    \label{fig:audio_attention_per_outputs}
\end{figure}

\textbf{Attention aggregation behavior analysis.}
Fig.~\ref{fig:audio_attention_per_outputs} visualizes the block-level attention distribution at Layer 14 during output generation. 
For each generated token, the attention weights are aggregated over four semantic input blocks: \textit{assistant input}, \textit{audio}, \textit{question}, and \textit{previous output}.
We observe that when the model generates tokens corresponding to temporal information (e.g., the numerical timestamps), the attention weight allocated to the \textit{audio} block significantly increases. 
This indicates that the model actively consults the audio representation when producing timestamp values.
This behavior supports the central intuition behind our framework: \emph{Grounding is Listening} as in Sec. \ref{subsec:grounding_is_listening}. 
When timestamp alignment is learned through STA, the model allocates more attention to audio tokens during the generation of temporal references, suggesting that timestamp prediction is grounded in actual acoustic evidence rather than language priors.
This qualitative observation complements our quantitative results in Section~\ref{sec:experiments}, which show that timestamp grounding improves both reasoning consistency and audio verification behavior.

\begin{figure*}[t]
    \centering

    \begin{subfigure}{0.45\linewidth}
        \centering
        \includegraphics[width=\linewidth]{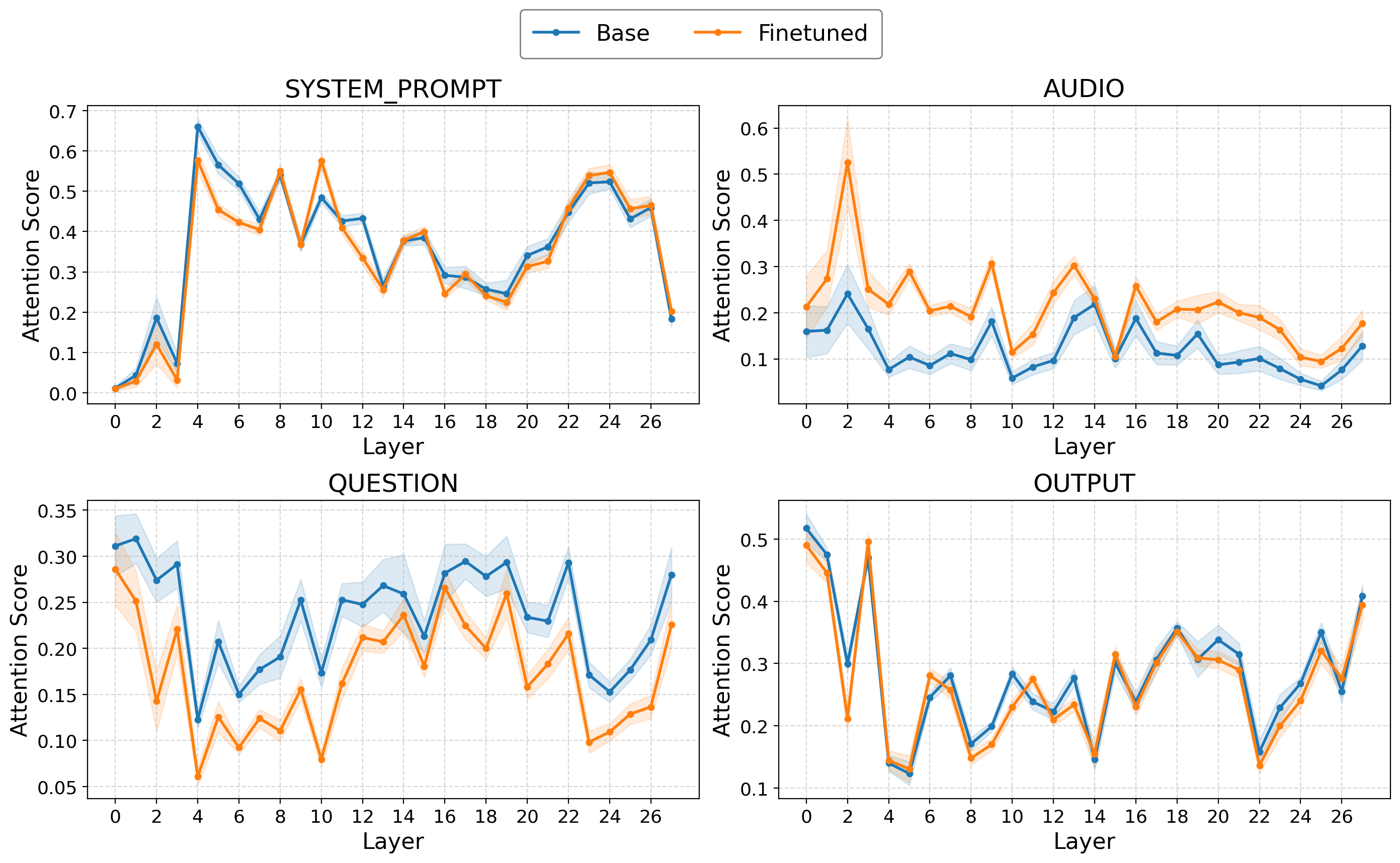}
        \caption{}
    \end{subfigure}
    \hfill
    \begin{subfigure}{0.45\linewidth}
        \centering
        \includegraphics[width=\linewidth]{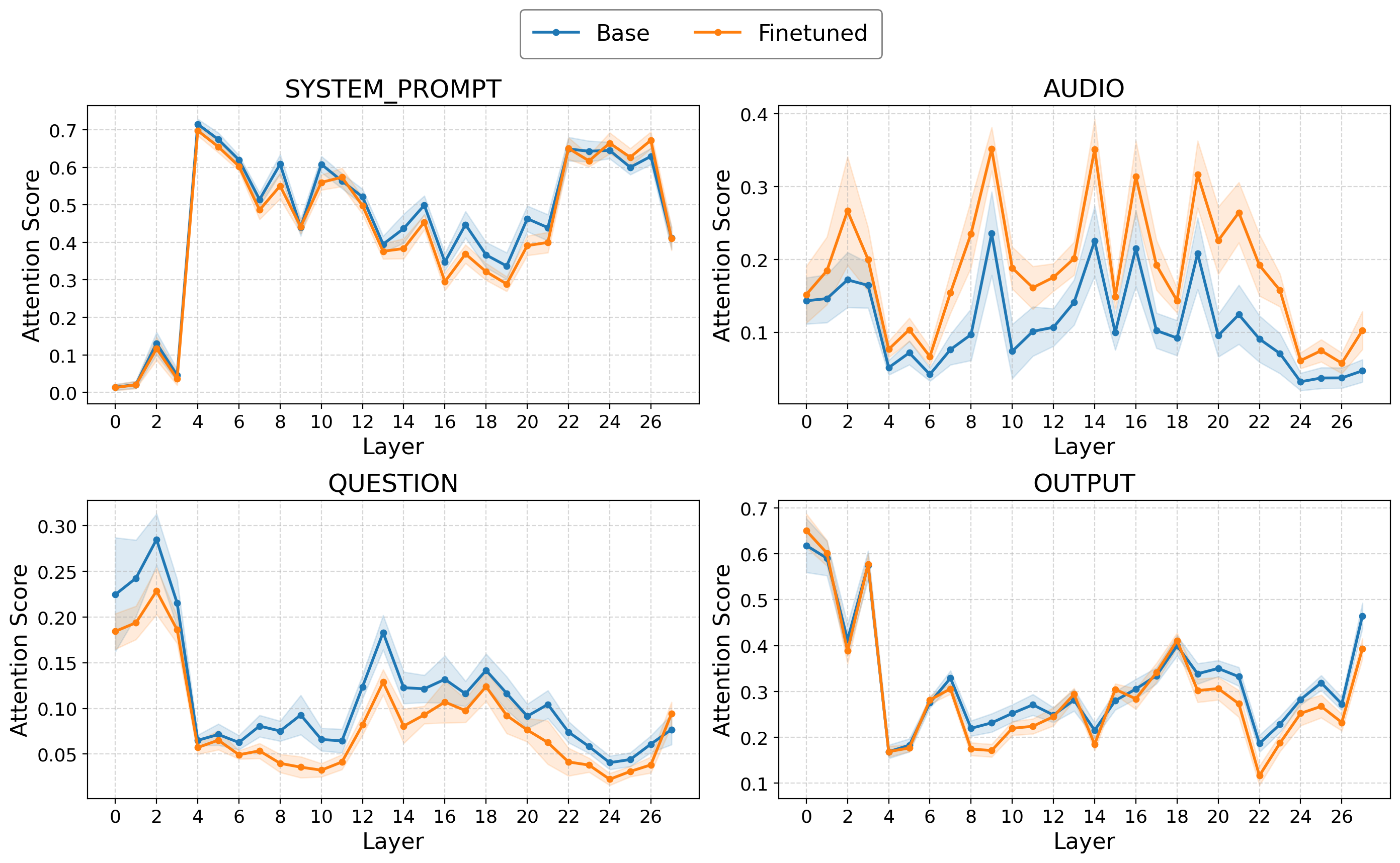}
        \caption{}
    \end{subfigure}

    \caption{
    \textbf{Layer-wise attention aggregation during timestamp generation across two LALM architectures.}
    (a) Audio Flamingo 3 and (b) Qwen2.5-Omni.
    Despite architectural differences, both models exhibit a similar shift toward increased reliance on acoustic representations during timestamp generation, supporting the tendency discussed in the main text.
    }
    \label{fig:layer_attention_tszone}
\end{figure*}

\subsection{Layer-wise Attention Statistics for Timestamp Generation}
\label{sec:appendix_attention_statistics}

To further examine how timestamp alignment influences the model's listening behavior, we analyze the layer-wise attention aggregation during timestamp generation discussed in Sec.~\ref{subsec:grounding_is_listening} and Fig.~\ref{fig:audio_attention_layers}. 
In this analysis, we report the attention allocated to entire semantic input blocks (\textit{assistant input}, \textit{audio}, \textit{question}, and \textit{self-referential outputs}) while the model generates timestamp tokens.
For each layer, we aggregate the attention scores across all output tokens associated with timestamp prediction.

Fig.~\ref{fig:layer_attention_tszone} presents the resulting statistics for two different LALMs: Audio Flamingo 3 and Qwen2.5-Omni.
Across both architectures, timestamp alignment consistently increases the amount of attention allocated to the \textit{audio} block across multiple layers.
Although these models differ substantially in their architectural design, we observe a similar shift toward increased reliance on acoustic representations during timestamp generation.
In addition, the increase in audio attention is not uniformly distributed across layers.
Instead, certain layers exhibit more pronounced changes, suggesting the presence of \textit{listening-sensitive layers} where acoustic evidence is selectively amplified.
A similar tendency is observable in Audio Flamingo 3 as well, indicating that such behavior is not specific to a single model.

Overall, these results provide complementary empirical evidence for our central claim that \textbf{grounding is listening}.
Beyond the examples shown in the main text, the same tendency is consistently observed across different architectures and layers: when timestamp alignment is introduced, models increasingly attend to audio tokens during temporal reasoning, suggesting that timestamp prediction becomes more tightly grounded in acoustic evidence.


\section{Additional Experimental Settings}
\label{section:B}

\subsection{Prompt Details}

\subsubsection{Stage 1: Supervised Timestamp Alignment for Qwen2.5-Omni}
\begin{tcolorbox}[enhanced, colback=gray!10!white, colframe=gray!40, arc=2mm, boxrule=0.6pt]
\small
\textbf{Q:} What is the timestamp of the following segment? [\emph{given sentence}] \\
\textbf{A:} The segment [\emph{given sentence}] starts at $t_s$ seconds and ends at $t_e$ seconds.
\end{tcolorbox}

\subsubsection{Stage 1: Supervised Timestamp Alignment for Audio Flamingo 3}
\begin{tcolorbox}[enhanced, colback=gray!10!white, colframe=gray!40, arc=2mm, boxrule=0.6pt]
\small
\textbf{Q:} What is the timestamp of the following segment? [\emph{given sentence}] \\
Provide both the start and end timestamps of this segment in seconds with 2 decimal places. \\ 
Format: The segment starts at X.XX seconds and ends at Y.YY seconds.

\textbf{A:} The segment starts at $t_s$ seconds and ends at $t_e$ seconds.
\end{tcolorbox}

\subsubsection{Stage 2: Timestamp-Grounded Reasoning Instruction}
\begin{tcolorbox}[enhanced, colback=gray!10!white, colframe=gray!40, arc=2mm, boxrule=0.6pt]
\small
When answering, you must first provide reasoning grounded in the audio content using explicit timestamps. \\ Start your response exactly with: 
\textit{``To determine the best description, let's analyze the audio content and the given timestamps:''}
Then, after the reasoning, provide the final answer.
\end{tcolorbox}

\subsection{Training Implementation Details}

\begin{table}[h]
\vspace{-4mm}
\footnotesize
\renewcommand{\arraystretch}{0.95}
\centering
\begin{minipage}{0.45\linewidth}
\centering
\caption{Training hyperparameters for STA supervised fine-tuning.}
\label{tab:training_hyperparameters}
\begin{tabular}{lc}
\toprule
\textbf{Hyperparameter} & \textbf{Value} \\
\midrule
Training Type & Full Fine-tuning \\
Precision & BF16 \\
Learning Rate & $2\times10^{-5}$ \\
Warmup Ratio & $0.03$ \\
Optimizer & AdamW \\
Epochs & $1$ \\
Number of GPUs & $4$ \\
Batch Size (per GPU) & $4$ \\
Gradient Accumulation & $8$ \\
LLM Parameters & Frozen \\
Audio Aligner & Trainable \\
\bottomrule
\end{tabular}
\end{minipage}
\hfill
\begin{minipage}{0.45\linewidth}
\centering
\caption{Training hyperparameters for GRPO-based reinforcement learning fine-tuning.}
\label{tab:grpo_training}
\begin{tabular}{lc}
\toprule
\textbf{Hyperparameter} & \textbf{Value} \\
\midrule
Training Type & LoRA \\
Precision & BF16 \\
Learning Rate & $5\times10^{-6}$ \\
Warmup Ratio & $0.03$ \\
Epochs & $1$ \\
Number of GPUs & $8$ \\
Batch Size (per GPU) & $8$ \\
Gradient Accumulation & $4$ \\
Number of Generations & $8$ \\
Temperature & $0.7$ \\
Top-$p$ & $0.9$ \\
KL Coefficient ($\beta$) & $0.04$ \\
Clipping Parameter ($\epsilon$) & $0.2$ \\
LLM Parameters & Trainable \\
Audio Aligner & Frozen \\
\bottomrule
\end{tabular}
\end{minipage}

\end{table}


\section{Additional Methodological Details}
\label{section:C}

\subsection{Problem Setup}
The objective of our work is to improve the reasoning capabilities of
LALMs through explicit temporal grounding.
We consider reasoning tasks defined by a dataset $\mathcal{D}$ of problem instances
$(A, q, a^{*})$, where $A$ denotes the speech audio input, $q$ is a natural language
query about the audio, and $a^{*}$ is the correct, verifiable answer.
The goal is to train a timestamp-grounded audio-language policy $\pi_{\theta}$,
parameterized by $\theta$, that produces a reasoning trace $\tau$ together with a
final answer $a$.
We define $\tau = \{(s_t, \Delta_t)\}_{t=1}^{T}$, where $s_t$ denotes the
$t$-th textual reasoning step and $\Delta_t \subset \mathbb{R}_{\ge 0}$ specifies the
corresponding temporal grounding region on the audio timeline.
The policy factorizes autoregressively as:
\begin{equation}
   \pi_{\theta}(\tau \mid A, q) =
   \left (\prod_{t=1}^{T}\pi_{\theta} \left(s_{t} \mid A, q, s_{<t}\right)\right)
   \cdot
   \pi_{\theta} \left ( a \mid A, q, s_{\le T}\right )
\end{equation}
where each reasoning step $s_t$ may include explicit temporal references $\Delta_t$ that anchor the linguistic reasoning to localized acoustic evidence within $A$.

\subsection{Group Relative Policy Optimization (GRPO)}

To optimize the timestamp-grounded audio-language policy, we adopt \emph{Group Relative Policy Optimization (GRPO)} \cite{shao2024deepseekmath}, a reinforcement learning framework that stabilizes policy learning by leveraging relative comparisons among multiple sampled trajectories.
Unlike standard policy gradient methods that rely on absolute reward values, GRPO evaluates each trajectory in the context of a group of alternatives, thereby reducing variance and mitigating reward scale sensitivity.

Given an input $(A, q)$, the policy $\pi_{\theta}$ samples a group of $G$ trajectories $\{\tau_i\}_{i=1}^{G}$.
Each trajectory is assigned a scalar reward $R_i$ according to task-specific verification criteria.
The normalized advantage for each trajectory is computed as
\begin{equation}
\tilde{R}_i =
\frac{R_i - \mathrm{mean}(R)}{\mathrm{std}(R)}
\end{equation}
where the mean and standard deviation are computed over the group $\{R_i\}_{i=1}^{G}$.
The policy is then optimized using an objective:
\begin{equation}
    \mathcal{J}_{\text{GRPO}}(\theta)=\mathbb{E}_{\mathcal{D}}\Bigg[\frac{1}{G}\sum_{i=1}^G\frac{1}{|\tau_i|}\sum_{t=1}^{|\tau_i|}\bigg(\min\left(\rho_{i,t}\tilde{R}_i,\\
    \text{clip}(\rho_{i,t},1-\epsilon,1+\epsilon)\tilde{R}_i \right)-\beta D_{KL}\left(\pi_\theta\, || \pi_{\text{ref}}\right)\bigg)\Bigg]
\end{equation}

where $\rho_{i,t}$ denotes the likelihood ratio between the current policy and the reference policy at step $t$, $\epsilon$ is the clipping threshold, and $\beta$ controls the strength of the KL regularization.

\subsection{RL Reward Motivation}
Recent advances in RL for reasoning have significantly improved the capabilities of LLMs in text-based domains \cite{shao2024deepseekmath, jaech2024openai}, enabling them to learn diverse reasoning strategies conditioned on the given context.
However, RL is known to primarily compose or amplify behaviors that are already latent in the sampling distribution of the base model rather than introducing entirely new reasoning capabilities \cite{yuedoes}.
In other words, RL can strengthen existing reasoning patterns but cannot reliably induce behaviors that are absent from the base model.

Unlike many recent reasoning approaches that rely on supervised fine-tuning with curated reasoning chains for cold-start initialization \cite{shao2024deepseekmath, jaech2024openai}, our model is not explicitly trained on timestamp-grounded reasoning chains prior to RL.
Instead, \emph{we verify that the base model after STA is already capable of producing timestamp-grounded reasoning behaviors through sampling.}
Based on this observation, we apply GRPO to further improve timestamp-grounded reasoning.
Nevertheless, as discussed in Sec.~\ref{subsec:grounding_behavior} and shown in Table~\ref{tab:behavior_analysis}, the existing reasoning chains still exhibit several limitations, including hallucinated timestamps, unstable grounding behavior, and imperfect benchmark performance.
These observations motivate the design of specialized reward functions that explicitly encourage temporally faithful reasoning.

\subsection{Behavior Analysis Metrics}
\label{sec:appendix_behavior_metrics}

We provide detailed descriptions of the metrics used in Sec.~\ref{subsec:grounding_behavior} and Table~\ref{tab:behavior_analysis}, which are designed to analyze how timestamp grounding affects the internal reasoning behavior of the model beyond end-task accuracy.

\textbf{Regions Explored} measures the number of distinct timestamped reasoning segments generated by the model.
Specifically, we count the number of unique temporal intervals $(t_s, t_e)$ referenced in the reasoning trace, together with their associated textual explanations.
This metric reflects how extensively the model explores different parts of the audio when constructing its reasoning.

\textbf{Audiology Verify} evaluates whether the model’s timestamped explanations are aligned with the actual audio content.
We implement the following verification pipeline:

\begin{enumerate}
    \item Parse predicted timestamps and their corresponding textual explanations from the model output.
    \item Extract audio segments based on each predicted interval $[t_s, t_e]$.
    \item Transcribe each cropped audio segment using a Whisper-based ASR model \cite{radford2023robust}.
    \item Compute text similarity between the transcription and the model-generated explanation.
\end{enumerate}

\textbf{Consistency} evaluates whether the model’s reasoning trace logically supports its final answer.
We employ an external LLM (Qwen3-32B \cite{yang2025qwen3}) as a judge to assess the alignment between the reasoning and the predicted answer.

Given a question, candidate choices, the model’s reasoning, and its final answer, the judge model outputs a binary verdict:
\begin{itemize}
    \item $1$: the reasoning supports the final answer (consistent)
    \item $0$: the reasoning does not support or contradicts the final answer
\end{itemize}

The judge is prompted to produce only a single binary decision without additional explanation.
Consistency is then computed as the average score across all evaluation samples.


\section{Limitations and Future Work}
\label{section:D}

While our approach improves temporal reasoning performance, several limitations remain and suggest directions for future research.

\textbf{Sparse reward signals.}
Our RL setup primarily provides final-answer-level rewards, while the proposed \textit{Timestamp Reasoning Compaction} implicitly introduces template-like structural guidance for timestamp-grounded reasoning.
Despite this sparse supervision, the model is already able to exhibit strong timestamp-grounded reasoning behavior, as evidenced by both quantitative results and human evaluation.
However, this setup may still lead to suboptimal credit assignment, where partially incorrect or incomplete reasoning chains receive positive rewards.
Introducing dense intermediate rewards that explicitly supervise both reasoning steps and grounding accuracy could further improve alignment between reasoning, grounding, and reward signals.

\textbf{Extended tool use and adaptive control.}
Our model is capable of locating and segmenting relevant audio regions, but it does not yet support richer forms of tool-augmented reasoning.
Incorporating additional capabilities such as audio highlighting, region comparison, or exploratory querying could further enhance compositional reasoning.
Future work may also explore adaptive control mechanisms that dynamically balance different actions during reasoning.

\textbf{Limited to the speech domain.}
Our current study focuses on the speech domain as mentioned in Sec.~\ref{sec:conclusion}.
Extending the framework to broader audio modalities, such as environmental sound and music, remains an important direction.
We expect that grounding mechanisms developed in this work can generalize to these domains, but this requires further investigation.


\newpage
\section{Qualitative Examples}
\label{section:E}

More examples can be found on our website \href{https://ijihoon98.github.io/TGSR/}{\textcolor{royalblue}{ijihoon98.github.io/TGSR/}}.

\begin{figure*}[h]
    \centering
    \vspace{40mm}
    \resizebox{0.8\linewidth}{!}{
    \includegraphics[width=\linewidth]{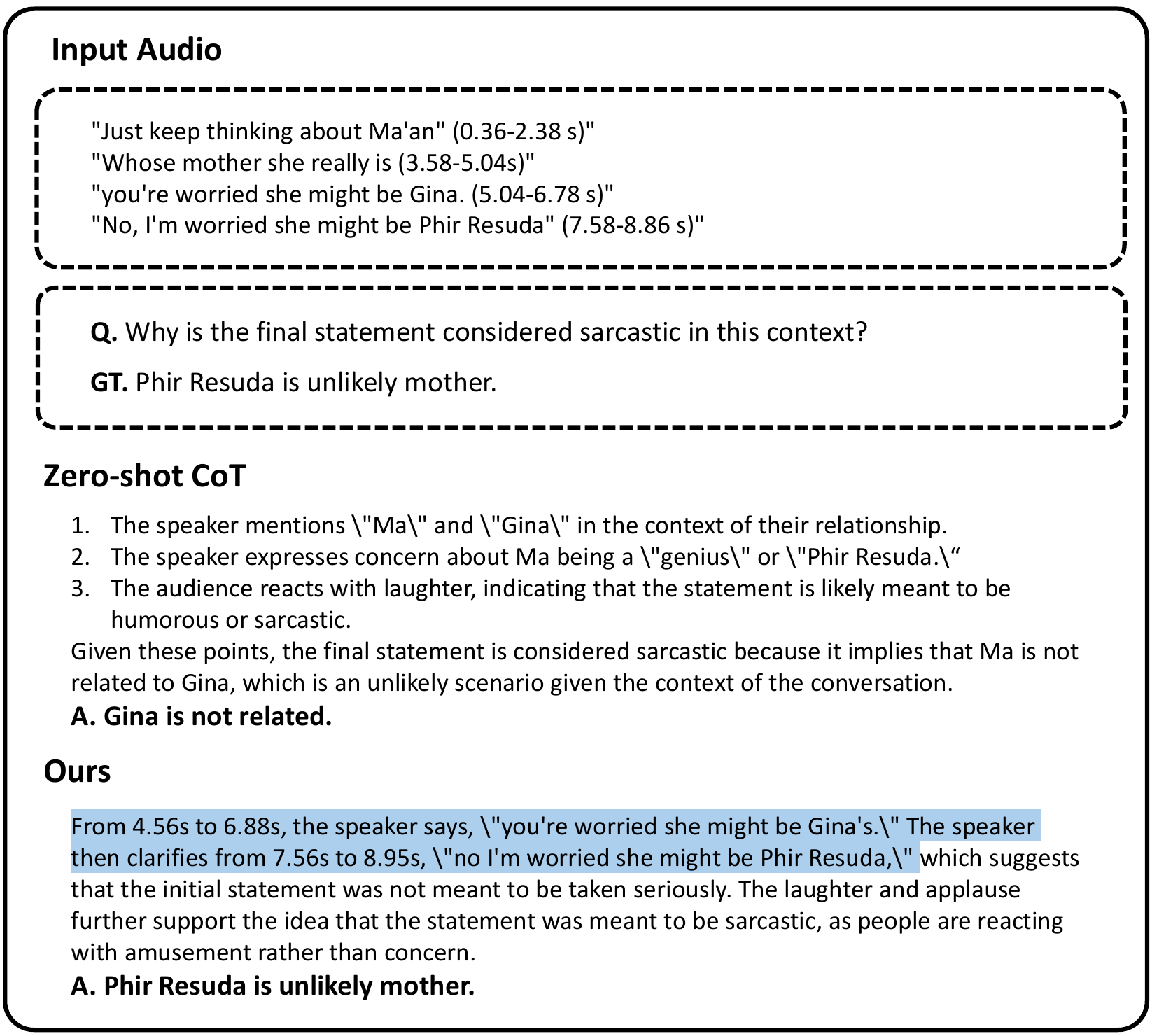}
    }
    \vspace{-2mm}
    \caption{
        \textbf{Comparison of reasoning traces between vanilla GRPO and our method on the MMAU dataset - 1.}
    }
    \vspace{-2mm}
\end{figure*}

\begin{figure*}[h]
    \centering
    \vspace{40mm}
    \resizebox{0.8\linewidth}{!}{
    \includegraphics[width=\linewidth]{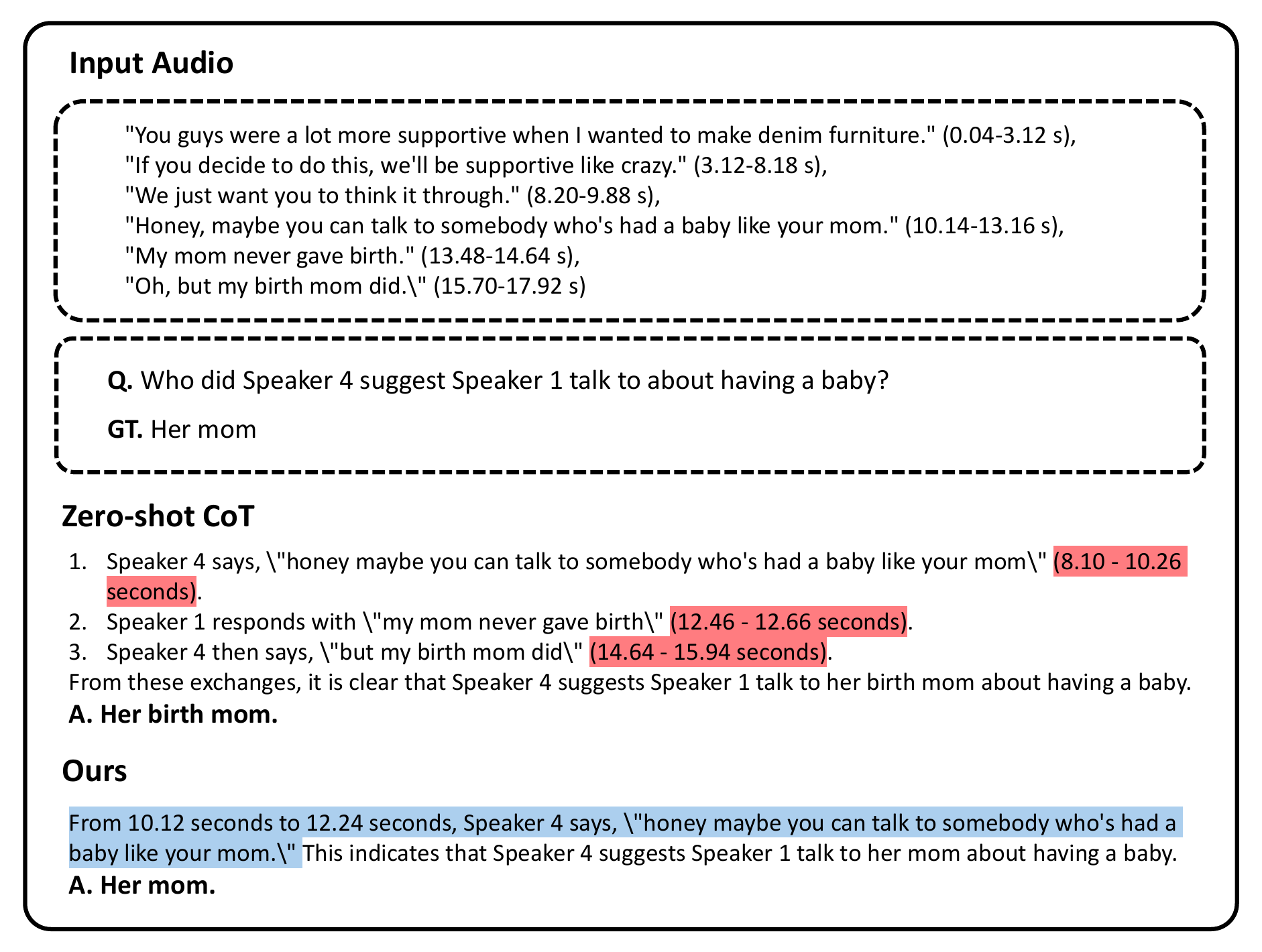}
    }
    \vspace{-2mm}
    \caption{
        \textbf{Comparison of reasoning traces between vanilla GRPO and our method on the MMAU dataset - 2.}
    }
    \vspace{-2mm}
\end{figure*}

\end{document}